\documentclass[prd,preprint,superscriptaddress,showpacs]{revtex4}
\begin{document}
\title{Solving Einstein Field Equations in Observational Coordinates 
with Cosmological Data Functions: Spherically Symmetric 
Universes with Cosmological Constant} 

\date{\today}

\author{M.E. Ara\'{u}jo}
\affiliation{ Departamento de F\'{\i}sica-Matem\'{a}tica, Instituto de F\'{\i}sica,\\
    Universidade Federal do Rio de Janeiro,\\
            21.945-970, Rio de Janeiro, R.J., Brazil}
\author{W.R. Stoeger}
\affiliation{ Vatican Observatory Research Group \\
      Steward Observatory, University of Arizona, \\
      Tucson, AZ 85721, USA}
\author{R.C. Arcuri}
\affiliation{ Departamento de F\'{\i}sica-Matem\'{a}tica, Instituto de F\'{\i}sica,\\
    Universidade Federal do Rio de Janeiro,\\
            21.945-970, Rio de Janeiro, R.J., Brazil}
\author{M.L. Bedran}
\affiliation{ Departamento de F\'{\i}sica,\\
        Universidade Federal de Juiz de Fora,\\
        36.036-330, Juiz de Fora, M.G., Brazil}

\begin{abstract}

Extending the approach developed by Ara\'ujo and Stoeger \cite{AS} and improved in 
Ara\'ujo {\it et al} \cite{ASR}, we have shown how to construct dust-filled
$\Lambda \neq 0$ Friedmann-Lema\^{\i}tre-Robertson-Walker (FLRW) cosmological models
from FLRW cosmological data on our past light cone. Apart from being of interest 
in its own right -- demonstrating how such data fully determines the models --
it is also illustrated in the flat case how the more general spherically symmetric (SS)
Einstein field equations can be integrated in observational coordinates with data fit to
FLRW forms arrayed on our past light cone, thus showing how such data determines a FLRW
universe -- which is not {\it a priori} obvious. It is also shown how to integrate these exact 
SS equations, in cases where the data are not FLRW, and the space-time is not known to be flat. 
It is essential for both flat and non-flat cases to have data giving the maximum of the observer
area (angular-diameter) distance, and the redshift $z_{max}$ at which 
that occurs. This enables the determination of the vacuum-energy density $\mu_{\Lambda}$, which 
would otherwise remain undetermined. \\

 \end{abstract}


\maketitle

\section{Introduction}

The recent WMAP's results [see Spergel {\it et al}  \cite{WMAP} and references therein] strongly support 
the inflationary scenario and are consistent with a nearly flat
Friedmann-Lema\^{\i}tre-Robertson-Walker Universe with a cosmological 
constant $\Lambda$ and dust ($\Omega \approx 1$, with $\Omega_m \approx 0.27$
and $\Omega_{\Lambda} \approx 0.73$), and with an almost scale invariant
spectrum for 
the primordial perturbations. It is obviously very important to test and
confirm this result. One observationally sensitive theoretical approach to
doing so begins by establishing a more general framework than FLRW -- say a 
general perturbed spherically symmetric space-time -- and then using the data 
itself to determine the more specific model. Can we obtain perturbed FLRW
by doing this? 

In two papers Ara\'ujo and Stoeger \cite{AS} and Ara\'ujo, Roveda and Stoeger \cite{ASR}
demonstrated in detail how to solve exactly the spherically symmetric (SS)
Einstein field equations for dust in observational coordinates without assuming
FLRW and with cosmological data representing galaxy
redshifts, observer area distances and galaxy number counts as
functions of redshift. These data are given, not on a space-like
surface of constant time, but rather on our past light cone
$C^-(p_0)$, which is centered at our observational position $p_0$
``here and now'' on our world line $ {\cal C}$.  These results
demonstrate how cosmologically relevant astronomical data can be used
to determine the space-time structure of the universe -- the
cosmological model which best fits it. This has been the aim
of a series of papers going back to the Physics Reports paper by Ellis
{\it et al.} \cite{Ellis et al}. The motivation and history of this `` observational
cosmology (OC) program''  is summarized in Ara\'ujo and Stoeger \cite{AS}. All
these papers assumed that $\Lambda = 0$. In this paper we demonstrate how
this program may be carried out when $\Lambda \neq 0$. As a simple, and very
relevant example, we take a flat SS universe, and suppose that the redshift,
observer area-distance, and number-count data can be fit to FLRW functional
forms (these are very special forms the data must take, if the universe
is FLRW). Then we show how such data determines a {\it bona fide} FLRW
universe -- which is not {\it a priori} obvious. We then go on to indicate
how the solution can be obtained in the more general non-flat case, with no
constrains on the functional form of the data functions. In doing this for
$\Lambda
\neq 0$, we also need another necessary piece of data, the maximum of the
observer area distance, and the redshift at which it occurs. Without these
extra observables, we do not have enough independent data to determine the
model -- in particular to determine the extra parameter $\Lambda$.

The primary aim of OC program is to strengthen the connections
between astronomical observations and cosmological theory. We do this
by allowing observational data to determine the geometry of spacetime
as much as possible, {\it without} relying on {\it a priori}
assumptions more than is necessary or justified. Basically, we want to
find out not only how far our observable universe is from being
isotropic and spatially homogeneous (that, is describable by an
FLRW cosmological model) on various length scales, but also to give a 
dynamic account of those deviations (Stoeger {\it et al} \cite{OC III}).

By using observational coordinates, we can thus formulate
Einstein's equations in a way which reflects both the geodesic flow of the
cosmological fluid and the null geometry of $C^-(p_0)$, along which 
practically all of our information about the distant reaches of our 
universe comes to us -- in photons. In this formulation the field equations
split naturally into two sets, as can be easily seen: a set of equations
which can be solved on $C^-(p_0)$, that is on our past light cone, specified
by $w = w_0$, where $w$ is the observational time coordinate; and a second
set which evolves these solutions off $C^-(p_0)$ to other light cones into
the past or into the future. Solution to the first set is directly determined 
from the data, and those solutions constitute the ``initial conditions'' for the
solution of the second set. 

There are many reasons for investigating FRLW in observational
coordinates from this more general starting point. It is clear from the
Cosmic Background Radiation (CBR) anisotropies measured by the Cosmic
Backgroung Explorer (COBE) and by WMAP that the universe is not exactly
described by the FLRW models (this follows from the analysis by Sachs and
Wolfe \cite{SW}, and see Stoeger, Ara\'ujo and Gebbie \cite{SAG} for an analysis
related to the viewpoint of this paper). But on the largest scales its
deviations from FLRW are small. So on those scales the universe can be
described by an {\it almost} FLRW model. Therefore, our first step
towards a strictly observationally-based approach to this realistic 
model involves a complete understanding of the inner workings of the
integration procedure in observational coordinates for FLRW data.  

In this paper, for completeness, we review some aspects of the problem
of determining the solution of the exact spherically symmetric
Einstein equations for dust in observational coordinates 
and then integrate the field equations with FLRW data to obtain the FLRW
($k=0$ and $\Lambda \ne 0$) solution explicitly. 
We refer the reader to Ellis {\it et al} \cite{Ellis et al}, Kristian and Sachs
\cite{KS}, Ara\'ujo and Stoeger \cite{AS} and references therein for a complete
account of the philosophy and the foundations of the OC approach leading to
the integration of Einstein field equations in observational coordinates.

In the next section we define observational coordinates, write the
general spherically symmetric metric using them and present the very
important central conditions for the metric variables. Section \ref{sec:obspar}
summarizes the basic observational parameters we shall be using and
presents several key relationships among the metric variables.
Section \ref{sec:fieqs} presents the full set of field equations for the spherically
symmetric case, with dust and with $\Lambda \neq 0$. Section \ref{sec:intpro}  shows the
integration procedure for the flat case, with FLRW data. In Section \ref{sec:nonflat}, we
present the integration for the general, non-flat case, and in section \ref{sec:concl} we
briefly discuss our conclusions.

\section{ The Spherically Symmetric Metric in Observational Coordinates}

We are using observational coordinates (which were first suggested by
Temple \cite{Temple}). As described by Ellis {\it el al}  \cite{Ellis et al}  the observational
coordinates $x^i=\{w,y,\theta ,\phi \}$ are centered on the observer's
world line $C$ and defined in the following way: \\

\noindent
(i) $w$ is constant on each past light cone along $C$, with $u^a
\partial _a w > 0$ along $C$, where $u^a$ is the 4-velocity of matter
($u^au_a=-1$). In other words, each $w = constant$ specifies a past
light cone along $C$. Our past light cone is designated as $w =
w_0$. \\

\noindent
(ii) $y$ is the null radial coordinate. It measures distance down the null
geodesics -- with affine parameter $\nu$ -- generating each past light cone
centered on $C$. $y = 0$ on $C$ and $dy/d\nu > 0$ on each null cone -- so
that $y$ increases as one moves down a past light cone away from $C$. \\

\noindent
(iii) $\theta$ and $\phi$ are the latitude and longitude of 
observation, respectively -- spherical coordinates based on a
parallelly propagated orthonormal tetrad along $C$, and defined away
from $C$ by $k^a \partial _a \theta = k^a \partial _a \phi = 0$, where
$k^a$ is the past-directed wave vector of photons ($k^ak_a=0$). \\

\noindent
There are certain freedoms in the specification of these observational
coordinates. In $w$ there is the remaining freedom to specify $w$ along our
world line $C$. Once specified there it is fixed for all other world lines.
There is considerable freedom in the choice of $y$ -- there are a large
variety of possible choices for this coordinate -- the affine parameter, $z$%
, the area distance $C(w,y)$ itself. We normally choose $y$ to be comoving
with the fluid, that is $u^a\partial _ay=0$. Once we have made this choice,
there is still a little bit of freedom left in $y$, which we shall use below.
The remaining freedom in the $\theta $ and $\phi $ coordinates is a rigid
rotation at $one$ point on $C$. 

In observational coordinates the Spherically Symmetric metric takes
the general form:
\begin{equation}
ds^2=-A(w,y)^2dw^2+2A(w,y)B(w,y)dwdy+C(w,y)^2d\Omega ^2,  \label{oc}
\end{equation}
where we assume that $y$ is comoving with the fluid, so that the
fluid 4-velocity is $u^a=A^{-1}\delta _w^a$.

The remaining coordinate freedom which preserves the observational form of
the metric is a scaling of $w$ and of $y$:
\begin{equation}
w\rightarrow \tilde{w}=\tilde{w}(w)~,~~y\rightarrow\tilde{y}= \tilde{y}%
(y)~~~~\left({\frac{d\tilde{w}}{dw}}\neq 0 \neq {\frac{d\tilde{y} }{dy}}%
\right).  \label{wy}
\end{equation}

The first, as we mentioned above, corresponds to a freedom to choose $w$ as
any time parameter we wish along $C$, along our world line at $y=0$. This is
usually effected by choosing $A(w,0)$. The second corresponds to the freedom
to choose $y$ as any null distance parameter on an initial light cone --
typically our light cone at $w=w_0$. Then that choice is effectively dragged
onto other light cones by the fluid flow -- $y$ is comoving with the fluid
4-velocity, as we have already indicated. We shall use this freedom to
choose $y$ by setting:
\begin{equation}
A(w_0, y) = B(w_0, y).  \label{ab}
\end{equation}
We should carefully note here that setting $A(w, y) = B(w, y)$ off our
past light cone $w = w_0$ is too restrictive.

In general, these freedoms in $w$ and $y$ imply the metric scalings:
\begin{equation}
A\rightarrow\tilde{A}={\frac{dw}{d\tilde{w}}}A~,~~ B\rightarrow\tilde{B}={%
\frac{dy}{d\tilde{y}}}B.  \label{4scale}
\end{equation}

It is important to specify the central conditions for the metric variables $%
A(w, y)$, $B(w, y)$ and $C(w, y)$ in equation (\ref{oc}) -- that is, their proper
behavior as they approach $y = 0$. These are:
\begin{eqnarray}
{\rm as}\;\;y\rightarrow 0:\;\;\; &&A(w,y)\rightarrow A(w,0)\neq 0, 
\nonumber \\
&&B(w,y)\rightarrow B(w,0)\neq 0,  \nonumber \\
&&C(w,y)\rightarrow B(w,0)y = 0,  \label{cent} \\
&&C_y(w,y)\rightarrow B(w,0).  \nonumber
\end{eqnarray}

\section{\label{sec:obspar}The Basic Observational Quantities}

The basic observable quantities on $C$ are the following: \\

(i) Redshift. The redshift $z$ at time $w_0$ on $C$ for a commoving source a
null radial distance $y$ down $C^{-}(p_0)$ is given by
\begin{equation}
1+z={\frac{A(w_0,0)}{A(w_0,y)}}.  \label{z}
\end{equation}
This is just the observed redshift, which is directly determined by source
spectra, once they are corrected for the Doppler shift due to local motions. %
\\

(ii) Observer Area Distance. The observer area distance, often written as $%
r_0$, measured at time $w_0$ on $C$ for a source at a null radial distance $y$
is simply given by
\begin{equation}
r_0=C(w_0,y),
\end{equation}
provided the central condition (\ref{cent}), determining the relation
between $C(w,y)$ and $B(w,y)$ for small values of $y$, holds. This quantity
is also measurable as the luminosity distance $d_L$ because of the reciprocity
theorem of Etherington \cite{Etherington33} (see also Ellis \cite{Ellis et al}),\\ 

\begin{equation}
d_L = (1+z)^2 C(w_0, y).   \label{recth} \\
\end{equation}

(iii) The Maximum of Observer Area Distance. Generally speaking, $C(w_0, y)$
reaches a maximum $C_{max}$ for a relatively small redshift $z_{max}$ (Hellaby
 \cite{Hellaby}; see also Ellis and Tivon  \cite{ET} and Ara\'{u}jo and Stoeger  \cite{ASII}). At
$C_{max}$, of course, we have 

\begin{equation}
\frac{d C(w_0, z)}{d z} = \frac{d C(w_0, y)}{d y} = 0,\\
\end{equation}
further conditioned by
\begin{equation}
\frac{d^2 C(w_0, y)}{d z^2} < 0.\\
\end{equation}
Furthermore, of course, as we shall
review below, the data set will give us $y = y(z)$, from which we shall be
able to find $y_{max} = y_{max}(z_{max})$.  These $C_{max}$ and $z_{max}$
data provide additional independent information about the cosmology. Without
$C_{max}$ and $z_{max}$ we cannot constrain the value of $\Lambda$. 
 
(iv) Galaxy Number Counts. The number of galaxies counted by a central
observer out to a null radial distance $y$ is given by
\begin{equation}
N(y)=4\pi\int_0^y \mu(w_0,\tilde{y})m^{-1}B(w_0,\tilde{y})C(w_0,\tilde{y})^2
d\tilde{y},  \label{N}
\end{equation}
where $\mu$ is the mass-energy density and $m$ is the average galaxy mass.
Then the total energy density can be written as
\begin{equation}
\mu(w_0,y) = m\;n(w_0,y) = M_0(z)\;{\frac{dz}{dy}}\;{\frac{1}{B(w_0,y)}},
\label{mudef}
\end{equation}
where $n(w_0, y)$ is the number density of sources at $(w_0, y)$, and where
\begin{equation}
M_0 \equiv {\frac{m}{J}}\;{\frac{1}{d\Omega}}\;{\frac{1}{r_0^2}}\;{\frac{dN}{dz}}. \label{m0def}
\end{equation}
Here $d \Omega$ is the solid angle over which sources are counted, and
$J$ is the completeness of the galaxy count, that is, the fraction of
sources in the volume that are counted is $J$. The effects of dark
matter in biasing the galactic distribution may be incorporated via $m$ and/or
$J$ . In particular, strong biasing is needed if the number counts have
a fractal behaviour on local scales (Humphreys {\it et al} \cite{hmm}). In order
to effectively use number counts to constrain our cosmology, we shall also
need an adequate model of galaxy evolution. We shall not discuss this
important issue in this paper. But, fundamentally, it would give us
an expression for $m = m(z)$ in equations (\ref{mudef}) and (\ref{m0def}) above.

There are a number of other important quantities which we catalogue here for
completeness and for later reference. 

First, there are the two fundamental four-vectors in the problem, the fluid
four-velocity $u^a$ and the null vector $k^a$, which points down the
generators of past light cones. These are given in terms of the metric
variables as
\begin{equation}
u^a = A^{-1}\delta^a{}_w ~,~~ k^a = (AB)^{-1}\delta^a{}_y.  \label{uk}
\end{equation}

Then, the rate of expansion of the dust fluid is $3 H = \nabla_a u^a$, so that,
from the metric (1) we have:
\begin{equation}
H={\frac{1}{3A}}\left({\frac{\dot{B}}{B}}+2{\frac{\dot{C}}{C}}\right),
\label{h}
\end{equation}
where a ``dot'' indicates $\partial/\partial w$ and a ``prime'' indicates $%
\partial/\partial y$, which will be used later. For the central observer $H$
is precisely the Hubble expansion rate. In the homogeneous (FLRW) case, $H$ is
constant at each instant of time t. But in the general inhomogeneous case, $%
H $ varies with radial distance from $y = 0$ on $t = t_0$. From our central
conditions above (3), we find that the central behavior of $H$ is given by
\begin{equation}
{\rm as}\;\;y\rightarrow 0:\;\;\;H(w,y)\rightarrow {\frac{1}{A(w,0)}}{\frac{%
\dot{B}(w,0)}{B(w,0)}}=H(w,0).  \label{hcent}
\end{equation}
At any given instant $w = w_0$ along $y = 0$, this expression is just the
Hubble constant $H_0 \equiv H(w_0, 0) = A_0^{-1} B_0^{-1}(\dot{B})_0$ as
measured by the central observer. In the above we have also written $A_0 \equiv A(w_0, 0)$
and  $B_0 \equiv B(w_0, 0)$.

Finally, from the normalization condition for the fluid four-velocity, we
can immediately see that it can be given (in covariant vector form) as the
gradient of the proper time $t$ along the matter world lines: $u^a=-t,_a$.
It is also given by (\ref{oc}) and (\ref{uk}) as
\begin{equation}
u_a=g_{ab}u^b=-Aw_{,a}+By_{,a}.
\end{equation}
Comparing these two forms implies
\begin{equation}
dt=Adw-Bdy~~\Leftrightarrow~~A=t_w~,~~ B=-t_y,  \label{tw}
\end{equation}
which shows that the surfaces of simultaneity for the observer are given in
observational coordinates by $A dw = B dy$. The integrability condition of
equation (\ref{tw}) is simply then
\begin{equation}
A^{\prime}+\dot{B}=0.  \label{coneq}
\end{equation}

This turns out precisely to be the momentum conservation equation, which is
the key equation in the system and essential to finding a solution. 

\section{\label{sec:fieqs}The Spherically Symmetric Field Equations in 
Observational Coordinates}

Using the fluid-ray tetrad formulation of the Einstein's equations 
developed by Maartens \cite{m} and Stoeger {\it et al} \cite{fluid ray}, 
one obtains the Spherically Symmetric field equations in observational 
coordinates with $\Lambda \neq 0$ (see Stoeger {\it et al} \cite{OC III} for a
detailed derivation).
Besides the momentum conservation equation (\ref{coneq}), they are
as follows.

A set of two very simple fluid-ray tetrad time-derivative equations can be
quickly integrated to give:
\begin{widetext}
\begin{equation}
\mu_m(w,y)=\mu_{m_0}(y)\;{\frac {B(w_0,y)} {B(w,y)}}\; \frac {C^{2}(w_0,y)} {C^{2}(w,y)}; 
\end{equation}

\begin{equation}
\omega(w,y)=\biggl(\omega_0(y) + {\frac {\mu_{\Lambda}} {6} }\biggr) {\frac{C^3(w_0, y)}{C^3(w,y)}}
- {\frac {\mu_{\Lambda}} {6}} = {-{\frac{1}{{2C^2}}}+{\frac{
\dot{C}}{{AC}}}{\frac{C^{\prime}}{{BC}}}+{\frac{1}{2}}\biggl({\frac{
C^{\prime}}{{BC}}}\biggr)^2}, \label{omega} 
\end{equation}
\end{widetext}
where $\mu_m$ again is the relativistic mass-energy density of the dust, 
including dark matter, and $\omega_0$, that is $\omega$ specified on $w = w_0$,  is a quantity
closely related to $\mu_{m_0}$ (see equation (\ref{omegdef}) below). In 
deriving and solving these equations, and those below, we have used the
typical $\Lambda$ equation of state, $p_{\Lambda} = - \mu_{\Lambda},$
where $p_{\Lambda}$ and $\mu_{\Lambda} \equiv \frac{\Lambda}{8 \pi G}$ are the pressure and the 
energy density due to the cosmological constant. Both $\omega_0$ and
$\mu_0$ are specified by data on our past light cone, as we shall show.
$\mu_{\Lambda}$ will eventually be determined from the measurement of $C_{max}$
and $z_{max}$. Essentially $\omega$ is defined by the right-hand-side of
equation (\ref{omega}).

The fluid-ray tetrad radial equations are:
\begin{widetext}
\begin{eqnarray}
&{\frac{C^{\prime\prime}}{C}} = {\frac{C^{\prime}}{C}}{\biggl({\frac{
A^{\prime}}{A}} +{\frac{B^{\prime}}{B}}\biggr)} - {\frac{1}{2}}B^2\mu_m;
\label{nr} \\
&\biggl[ \bigl(\omega_0(y) + {\frac {\mu_{\Lambda}} {6} }\bigr)C^3(w_0, y)\biggr]^{\prime} = -{\frac{1}{2}}\mu_{m_0}\;
{B(w_0,y)}\;{C^{2}(w_0,y)}\;{\biggl({\frac{\dot {C}}{A}} +
 {\frac{C^{\prime}}{B}}\biggr)};  \label{omegap} \\
&{\frac{{\dot {C}}^{\prime}}{C}} = {\frac{{\dot B}}{B}}{\frac{C^{\prime}}{C}
} - \biggl(\omega + {\frac {\mu_{\Lambda}} {2}}\biggr)\; A\;B.  \label{prdot}
\end{eqnarray}
\end{widetext}
The remaining ``independent'' time-derivative equations given by the
fluid-ray tetrad formulation are:

\begin{eqnarray}
&&{\frac{{\ddot{C}}}C}={\frac{{\dot{C}}}C}{\frac{{\dot{A}}}A}+ \biggl(\omega + 
{\frac {\mu_{\Lambda}} {2}}\biggr) \;A^2;
\label{dotdot} \\
&&{\frac{{\ddot{B}}}B}={\frac{{\dot{B}}}B}{\frac{{\dot{A}}}A}-2\omega \;A^2-{%
\frac 12}\mu_{m} \;A^2 .  \label{bdd}
\end{eqnarray}
From equation (\ref{omegap}) we see that there is a naturally defined
``potential'' (see Stoeger {\it et al} \cite{OC III}) depending only on the radial
null coordinate $y$ -- since the left-hand-side depends only on $y$, the
right-hand-side can only depend on $y$:

\begin{equation}
F(y)\equiv {\frac{{N_{\star }}^{\prime }}{N^{\prime }}}={\frac{\dot{C}}A}+{%
\frac{C^{\prime }}B},  \label{f1}
\end{equation}
where $N_{\star }(y)$ is an arbitrary function, whose central behavior is
the same as that of number counts (Stoeger {\it et al} \cite{OC III}). Thus,
\begin{widetext}
\begin{equation}
\omega_0(y)= - {\frac {\mu_{\Lambda}} {6} }- {\frac{1}{2 C^3(w_0, y)}}\;\int{\mu_{m_0}(y)\;
{B(w_0,y)}\;{C^{2}(w_0,y)}\;F(y)\;dy}.  \label{omegdef}
\end{equation}
Connected with this relationship is equation (\ref{omega}), which we rewrite
as 
\begin{equation}
{\frac{{\dot C}}{C}}{\frac{C^{\prime}}{C}}+{\frac{A}{2B}}{\frac{{C^{\prime}}%
^2}{C^2}}-{\frac{AB}{2C^2}} = {\frac{AB}{C^3}\biggl[C_0^3\bigl(\omega_0 + \frac{\mu_{\Lambda}}{6}\bigr) - \frac{\mu_{\Lambda}}{6}{C^3}\biggr]},  \label{f2}
\end{equation}
where $C_0 \equiv C(w_0, y)$.

Stoeger {\it et al} \cite{OC III} and Maartens {\it et al} \cite{InhomUniv}
have shown that equations (\ref{f1}) and (\ref{f2}) can be transformed into
equations for $A$ and $B$, 
thus reducing the problem to determining $C$:
\begin{eqnarray}
&&A = {\frac{{\dot C} }{[F^2 -1 -2(\omega_0 + \mu_{\Lambda}/6) C_0^3/C
+ (\mu_{\Lambda}/3) C^2]^{1/2}}}  \label{aeq} \\
&&B = {\frac{C^{\prime} }{F \pm [F^2 -1 -2(\omega_0 + \mu_{\Lambda}/6)C_0^3/C
+ (\mu_{\Lambda}/3) C^2]^{1/2}}}.  \label{beq}
\end{eqnarray}
\end{widetext}
The Lema\^{\i}tre-Tolman-Bondi (LTB) exact solution (Lema\^{\i}tre \cite{Lemaitre}), Tolman \cite{Tolman}, Bondi \cite{Bondi}; and cf. Humphreys \cite{HT} and 
references therein) is
obtained by integration of (\ref{aeq}) along the matter flow
$y= $constant using (\ref{tw})
\begin{equation}
t-T(y)=\int \frac {dC}{[F^2 -1 -2(\omega_0 + \mu_{\Lambda}/6) C_0^3/C + (\mu_{\Lambda}/3) C^2]^{1/2}},  \label{bs}
\end{equation}
where $T(y)$ is arbitrary, provided we identify
\begin{equation}
F^2=1-kf^2,\;\;\;\;k=0,\pm 1.  \label{f}
\end{equation}
Here $f=f(y)$ is a function commonly used in describing LBT 
models in the 3 + 1 coordinates (Bonnor \cite{Bonnor}).  

\section{\label{sec:intpro}Integration with FLRW ($k=0$ and $\Lambda \neq 0$) Data}

In this section and the next we use a generalization (to incorporate the
 $C_{max}$ and
$z_{max}$ data) of the integration procedure described in detail in Ara\'ujo 
{\it et al} \cite{ASR}, which in turn is an improvement of the integration scheme
developed by Ara\'ujo and Stoeger \cite{AS}, to solve the above system of
SS field equations when $\Lambda \neq 0 $. First, we consider a concrete,
simplified, but very
relevant example. Suppose that we know that the universe is flat. Then 
$F(y) = 1$. This means that we need only the observer area distance, or the 
galaxy number counts -- not both. $F=1$ establishes a relation between these
data functions. Suppose also that, though we do not know that the universe
is FLRW, we find that our observer-area-distance and galaxy-number
count data can be fit -- or very closely approximated -- by the FLRW,
$\Lambda \neq 0$ observational relationships as functions of the redshift $z$.
For the flat case these are (Ara\'{u}jo and Stoeger \cite{ASII}): 
\begin{widetext}
\begin{eqnarray}
r_0(z)&=&{\frac {[Q(z)-1]}{\sqrt {\Omega_{\Lambda} {\sqrt 3}} H_0 [(1+\sqrt3)Q(z)-(1-\sqrt3)]}} 
\Biggl\{Cn^{-1} \biggl[ Q(z), \kappa \biggr] 
       -Cn^{-1} \biggl[ Q(0) , \kappa \biggr]\Biggr\}, \nonumber \\
Q(z) &\equiv& \frac{(1-\sqrt 3)+(1+z){\root 3 \of {{1\over \Omega_{\Lambda}}-1}}}{(1+\sqrt 3)+(1+z)
{\root 3 \of {{1\over \Omega_{\Lambda}}-1}}}, \nonumber \\
\Omega_{\Lambda} &\equiv&  {\frac {\Lambda} {3{H_0}^2}}, \nonumber \\
\kappa&=&\sqrt{\frac {2+\sqrt3} {4}}, \label{r0}
\end{eqnarray}
and 
\begin{equation}
M_0(z)= {\frac {\mu_{m_0}(1+z)^2} {A_0 H_0\sqrt{[\Omega_{\Lambda}+ (1+z)^3(1-\Omega_{\Lambda})]}}}. \label{m0}
\end{equation}
\end{widetext}
$Cn^{-1}(u, \kappa)$ is the inverse of the Jacobi elliptic function $Cn(u,
\kappa)$, where $\kappa$ is the modulus. Equation  (\ref{r0}) can clearly also be
written in terms of elliptic integrals of the first kind (Ara\'{u}jo and Stoeger
\cite{ASII}). \\

Equations  (\ref{r0}) and  (\ref{m0}) are the $\Lambda \neq 0$ analogues of the familiar
characteristic FLRW $r_0 = r_0(z)$ and $M_0 = M_0(z)$ relationships for
$\Lambda = 0$ (Ellis and Stoeger \cite{ES1987}; Stoeger, {\it et al.} \cite{OC III}): If the 
universe is FLRW and $\Lambda = 0$, $r_0(z)$ and $M_0(z)$ will have those
functional forms. Equations  (\ref{r0}) and  (\ref{m0}) -- or their elliptic-integral
equivalents -- are the corresponding characteristic  functional
forms for flat FLRW, $\Lambda \neq 0$ cases. As in the $\Lambda = 0$ cases,
however, it is not a trivial conclusion that data which satisfies
equations  (\ref{r0}) and  (\ref{m0})
implies an FLRW, $\Lambda \neq 0$ universe. This must be demonstrated -- and
can be demonstrated -- by using equations  (\ref{r0}) and  (\ref{m0}) as data functions to
solve the field equations and obtain an FLRW solution. This was done for the
flat $\Lambda = 0$ case by Ara\'{u}jo and Stoeger \cite{AS}. We now do the same
for these flat $\Lambda \neq 0$ cases. \\     

Because of the additional parameter ($\Lambda$) in the equations, though, in the
general (non-flat) case we need more data than supplied simply by equations
 (\ref{r0}) and  (\ref{m0}). When we know the universe is flat, this extra data can be
supplied by equation (\ref{m0}), which will then automatically enable us to
calculate $\Omega_{\Lambda}$ -- from the flatness condition. But also on $w =
w_0$ we can observationally determine where $C_0 \equiv r_0$ 
reaches its maximum value $C_{0max}$ and the redshift $z_{max}$ at which this
occurs. These measurements provide the needed extra data in the general case,
and can also be used, instead of the data in equation (\ref{m0}) in the flat
case. From equation (\ref{r0}), we can
immediately determine the equation for this maximum redshift, which will be

\begin{equation}
dr_0(z)/dz = 0.  \label{drdz} \\
\end{equation}

Plugging the observationally determined values of $z_{max}$ into this equation,
we obtain a unique relationship between $z_{max}$ and $\Omega_{\Lambda}$ 
(Ara\'{u}jo and Stoeger  \cite{ASII}), since $H_0$ cancels out of equation  (\ref{drdz}). Using
this relationship along with $C_{0max}$ in equation (\ref{r0}) will also 
determine $H_0$. The precise interpretation and definition of these parameters,
e.g. that $\Omega_{\Lambda}$ -- depending not only on $\Lambda$ but also on
$H_0$, the FLRW Hubble parameter -- is the density parameter for $\Lambda$, is
in reference to a supposed FLRW universe ($H_0$, and therefore
$\Omega_{\Lambda}$, cannot
be unambiguously defined in a general exactly spherically symetric -- also
often referred to as a Lema\^{i}tre-Tolman-Bondi (LTB) -- universe; the
definition of the rate of expansion given in equation (\ref{h}) is not the only
one that could be chosen, or measured). That
assumption is validated by continuing with the integration and showing that a
universe with such data is indeed FLRW. \\
 
Of course, this is a somewhat contrived case, since we would
not attempt to fit the data to such a functional form (\ref{r0}) unless we
already suspected that the universe may be FLRW, and that therefore
$\Omega_{\Lambda}$ and $H_0$, the
FLRW density parameter and the Hubble parameter at $w = w_0$ and $y = 0$,
can be defined in terms of an FLRW model. But besides being a very
relevant simple case, it serves to illustrate the integration procedure with
definite meaningful data input functions. \\ 

Solving the null Raychaudhuri equation (\ref{nr}) on $w = w_0$ with this data
(see Stoeger
{\it et al}  \cite{OC III}, and also Ara\'{u}jo and Stoeger \cite{OC III}) yields the following
relation between
redshift and the null coordinate $y$:
\begin{eqnarray}
1+z &=& \frac {(1-\sqrt 3)-(1+\sqrt 3)Cn(Ly+\sigma)}{ {\root 3 \of {{1\over \Omega_{\Lambda}}-1}} \Big [Cn(Ly+\sigma)-1\Big ]}, \nonumber \\
L&\equiv&-{\sqrt {\Omega_{\Lambda} {\sqrt 3}}}A_{0}H_{0} \root 3 \of {{1\over \Omega_{\Lambda}}-1} = -{1 \over w_0}, \nonumber \\
\sigma&\equiv&Cn^{-1} \biggl[ Q(0) , \kappa \biggr]. 
\label{zy}
\end{eqnarray}

Using equation (\ref{z}) we can now write $A(w_0,y)$ as a function of y as
\begin{widetext}
\begin{equation}
A(w_0,y) = -A_0\Biggl(\root 3 \of {{1\over \Omega_{\Lambda}}-1}\Biggr)\frac{\big[1-Cn(Ly+\sigma)\big]}{\big[(1-\sqrt 3)-(1+\sqrt 3)Cn(Ly+\sigma)\big]} \label{Aw0y}.
\end{equation}
$A_0 \equiv A(w_0, 0)$ is a constant scaling factor which can be chosen
arbitrarily. \\

If we wish to continue using our assumption that this will be an FLRW universe,
then the observer area distance $C(w_0,y)$ as a function of $y$ can be
written as 
\begin{equation}
C(w_0,y) = -A_0 y\Biggl(\root 3 \of {{1\over \Omega_{\Lambda}}-1}\Biggr)\frac{\big[1-Cn(Ly+\sigma)\big]}{\big[(1-\sqrt 3)-(1+\sqrt 3)Cn(Ly+\sigma)\big]}. 
\label{oad}
\end{equation}
\end{widetext}
If we don't wish to assume FLRW here, we can still find $C(w_0, y)$ by using
equation (\ref{r0}) in conjunction the result given in equation (\ref{zy}). \\

Furthermore, we can clearly now determine what $y_{max}$ is, corresponding to
$z_{max}$, from equation (\ref{zy}). \\ 

Now we begin to move our solution off our past light cone, $w=w_0$.
Since $y$ is chosen to be a comoving radial coordinate, the
functional dependence of $A(w,y)$ with respect to $y$ can not change
as we move off our light cone. We have already mentioned the freedom 
we have, temporally setting aside central-condition considerations, 
to rescale the time coordinate $w$, which is affected by choosing $A(w,0)$. 
Therefore, this freedom effectively corresponds to choosing the
functional dependence of $A(w,y)$ with respect to $w$ in any way we like, 
constrained only by the form of $A(w_0,y)$ [later this choice may have to 
be adjusted to satisfy all the central conditions, those on $C(w,y)$ and $B(w,y)$]. In our
expression for $A(w_0, y)$ is hidden an implicit dependence on $w$.
We need to extract that dependence and make it explicit, so that we
can then determine the general dependence of $A$ on $w$ and proceed
with the integration. In general, this is not simply achieved by 
replacing $w_0$ with $w$, because -- besides the $w_0$ dependence arising 
from setting $w = w_0$ when we write equation (\ref{z}) -- there is another 
part of the $w_0$ dependence which derives from integration constants of the
null Raychaudhuri equation and remains through the entire problem. 

At this point, accordingly to what we just pointed out, we arbitrarily set the $w$ dependence 
for $A$ and proceed with the integration. The next step is then the solution of equations
(\ref{coneq}) and (\ref{aeq}) to determine $B$ and $C$ respectively.
Their formal general solutions are:

\begin{equation}
\label{bint}
B= - \int{A^{\prime} dw} + l(y),
\end{equation}
where $l(y)$ is determined from the condition $A(w_0,y)=B(w_0,y)$, and

\begin{equation}
\label{cint}
C= \left [ -\Biggl(\frac{6\omega_0}{\mu_{\Lambda}}+1\Biggr){C_0}^3 {\sinh}^2 \Biggl(
{\root \of {3\mu_{\Lambda} \over 4}} \int{A dw}\Biggr) + h(y) \right ]^{1/3},
\end{equation}
where $h(y)$ is determined from the data $r_0=C(w_0,y)$. Since we know
$C(w_0,y)$ from equation (\ref{oad}), or from equation (\ref{r0}) with
equation (\ref{zy}), $\omega_0(y)$ is obtained from equation (\ref{beq}) and
is given by
\begin{equation}
\omega_0= - {\frac{1}{2C^2}} \Biggl [ 1 -{\frac{{(1+z)C^{\prime}}}{{A_0}}}\Biggr ]^2 \label{omrelc},
\end{equation}
where we have used (\ref{ab}) and (\ref{z}) to write
\begin{equation}
B(w_0,y)=A(w_0,0)/\left [1+z(y)\right ]. \label{baz}
\end{equation}
Here $z(y)$ is given by equation (\ref{zy}). When we do not know that
the universe is flat, we must, of course, first determine $\dot{C}(w_0, y)$,
in order to determine $F(y)$ from equation (\ref{f1}). This can be easily done,
as explained in Maartens, {\it et al}  \cite{InhomUniv}, in Ara\'{u}jo and Stoeger  \cite{ASII}, and
in the next section. \\

$B(w,y)$ and $C(w,y)$ are then determined by integrating
equations (\ref{coneq}) and (\ref{aeq}) with respect to $w$. 
$B(w,y)$ and $C(w,y)$ are further constrained, as discussed above, 
by the fact that they have to satisfy the central
conditions (\ref{cent}). Now, it is clear from an examination of these
equations that unless $A(w,y)$ has a very specific functional
dependence on $w$ the resulting solutions $B(w,y)$ and $C(w,y)$ will
not satisfy the central conditions. That implies that, although we can
find solutions to the field equations, it does not guarantee that the
null surface on which we assume we have the data is a past light cone
of our world line (Ellis {\it et al}  \cite{Ellis et al}). So we conclude that given the
fulfilment of the following conditions:
{\it

(1) $A(w_0,y)$ is determined  by the data and the central conditions;

(2) The coordinate $y$ is choosen to be a comoving radial coordinate;

(3) The central conditions }(\ref{cent});

\noindent {we can remove the freedom of rescaling the time coordinate $w$ and
completely determine
$A(w,y)$. Thus, all the coordinate freedom in $y$ and $w$ has been used up at 
this stage.}

Therefore, following the above analysis we find that the appropriate form for $A(w,y)$ is

\begin{widetext}
\begin{equation}
A(w,y) = -A_0\Biggl(\root 3 \of {{1\over \Omega_{\Lambda}}-1}\Biggr)
\frac{1-Cn\Bigl[(\sigma-1)+\frac{w-y}{w_0}\Bigr]} {\big(1-\sqrt 3)-(1+\sqrt 3)Cn\Bigl[(\sigma-1)+ \frac{w-y}{w_0} \Bigr]}, \label{Awy}
\end{equation}
where $z(y)$ is given by equation (\ref{zy}). Now, observing that 
\begin{equation}
{dz\over dy}=A_0H_0\sqrt{ \Omega_{\Lambda} + (1+z)^3(1-\Omega_{\Lambda})},
\end{equation}
we substitute $A^{\prime}(w,y)$ and $A(w,y)$ into equations (\ref{bint}) and (\ref{cint}) 
and determine the arbitrary functions of $y$ that arise from these integrations 
by the conditions $B(w_0,y)=A(w_0,y)$ and $C(w_0,y)=r_0(y)$ respectively. Thus,

\begin{equation}
B(w,y)= A(w,y) = -A_0\Biggl(\root 3 \of {{1\over \Omega_{\Lambda}}-1}\Biggr)
\frac{1-Cn\Bigl[(\sigma-1)+\frac{w-y}{w_0}\Bigr]} {\big(1-\sqrt 3)-(1+\sqrt 3)Cn\Bigl[(\sigma-1)+ \frac{w-y}{w_0} \Bigr]}, \label{Bwy}
\end{equation}
and
\begin{equation}
C(w,y)= A(w,y) y = -A_0 y\Biggl(\root 3 \of {{1\over \Omega_{\Lambda}}-1}\Biggr)
\frac{1-Cn\Bigl[(\sigma-1)+\frac{w-y}{w_0}\Bigr]} {\big(1-\sqrt 3)-(1+\sqrt 3)Cn\Bigl[(\sigma-1)+ \frac{w-y}{w_0} \Bigr]}, \label{Cwy}
\end{equation}
\end{widetext}
which are the FLRW form of the solutions for $\Lambda \neq 0$ in observational
coordinates. One can easily check (after some algebra) that the central conditions ({\ref{cent}}) are all satisfied, 
which in turn guarantees that the null surface on which we assume we have the data is indeed a past light cone of our world line.

\section{\label{sec:nonflat}The General Solution in the Non-Flat Case}

We now outline the integration procedure in the case where
we do not know whether the universe is flat or not, and where data gives us 
redshifts $z$, observer area distances (angular-diameter distances) $r_0(z)$, 
``mass source densities'' $M_0(z)$ which cannot be fit by the FLRW functional
form, and the angular-distance maximum
$C_{max}(w_0, z)$ at $z_{max}$. It is important to specify the latter, because,
as we have already emphasized, without them, we do not have enough information
to determine all the parameters of the space-time in the $\Lambda \neq 0$ case.
For instance, although we can determine $C(w_0, z)$ with good precision
(by obtaining luminosity distances $d_L$ and employing the reciprocity theorem,
equation  (\ref{recth})) out to relatively high redshifts, at present we do not yet have
reliable data deep enough to determine $C_{max}$ and $z_{max}$. But this has
just recently become possible with precise space-telescope distance
measurements of distance for supernovae Ia. \\

In pursuing the general integration with these data, we use the framework
and the intermediate results we have presented in Section \ref{sec:fieqs}. Obviously, one
of the key steps we must take now is the determination of the ``potential''
$F(y)$, given by equation (\ref{f1}). This was done in a similar way for 
$\Lambda = 0$ by Ara\'{u}jo and Stoeger  \cite{AS}, as indicated above. This means
we need to determine $C^{\prime}(w_0, y)$ and $\dot{C}(w_0, y)$, which we
now write as $C_0^{\prime}$ and $\dot{C}_0$, respectively. We also need
$A(w_0, y).$ We remember, too, that at on $w = w_0$ we have chosen
$B(w_0, y) = A(w_0, y)$, which we have the freedom to do. \\

Clearly, $C_0^{\prime}$ can be determined from the $r_0(z) \equiv C(w_0,z)$
data, through fitting, along with the solution of the null Raychaudhuri 
equation (\ref{nr}), as indicated in Section \ref{sec:intpro}, to obtain $z = z(y)$. $A(w_0, y)$,
too, is obtained from redshift data along with this same $z(y)$ result. 
$\dot{C}_0$ is somewhat more difficult to determine. But the procedure is
straight-forward.\\

We determine $\dot{C}_0$ by solving equation (\ref{prdot}) for it on
$w = w_0$. Using equations (\ref{ab}) and (\ref{coneq}), we can write this now as:
\begin{equation}
\frac{\dot{C}_0^{\prime}(y)}{C_0(y)}= -\frac{A_0^{\prime}(y)C_0^{\prime}(y)}
{A_0(y)C_0(y)} - A_0^2(y)(\omega_0(y) + \mu_{\Lambda}/2). \label{cdpz}
\end{equation}
But, from equation (\ref{omega}) we can write $\omega_0(y)$ in terms of $C_0(y)$,
$C_0^{\prime}(y)$, and $\dot{C}_0(y)$. So equation {\ref{cdpz}} becomes:
\begin{widetext}
\begin{equation}
\dot{C}_0^{\prime}(y) + \frac{C_0^{\prime}(y)\dot{C}_0(y)}{C_0(y)} =
\frac{A_0^2(y)}{2C_0(y)} -\frac{A_0^{\prime}(y)}{A_0(y)}C_0^{\prime}(y) -
\frac{(C_0^{\prime}(y))^2}{2C_0(y)} + \frac{A_0^2(y)C_0(y)}{2}\mu_{\Lambda}.
\label{cdpzfin}
\end{equation}
\end{widetext}
This is a linear differential equation for $\dot{C}_0(y)$, where from data
we know everything on our past light cone, $w = w_0$, (once the null
Raychaudhuri equation (\ref{nr}) has been solved) except $\dot{C}_0(y)$ itself and
$\mu_{\Lambda}$, which is a constant that can be carried along and determined
subsequently from $C(w_0, z_{max})$ and $z_{max}$ measurements (see below). 
Thus, we can easily solve equation (\ref{cdpzfin}) for $\dot{C}_0(y)$, which
will also depend on the unknown constant $\mu_{\Lambda}$. \\

However, introducing this result back into equation (\ref{cdpzfin}), and
evaluating it at $y_{max}$, which corresponds to $z_{max}$, we have simply
\begin{equation}
\dot{C}_0^{\prime}(y_{max}) = \frac{A_0^2(y_{max})}{2C_0(y_{max})} -
        \frac{A_0^2(y_{max}) C_0(y_{max})}{2}\mu_{\Lambda}, \label{cdpmax}
\end{equation}
where $\dot{C}_0^{\prime}(y_{max})$, as we have already emphasized, also
depends on the unknown $\mu_{\Lambda}$. Since everything else is now known,
equation (\ref{cdpmax}) is now an algebraic equation for
$\mu_{\Lambda}(y_{max})$, or equivalently for $\mu_{\Lambda}(z_{max})$. 
Obviously, we could have simply worked out this result in terms of $z_{max}$ to
begin with. \\

With this determination of $\mu_{\Lambda}$, we know $\dot{C}_0(y)$ completely,
and can now determine $F(y)$ from equation (\ref{f1}). From there on, we can follow
the solution off $w = w_0$ for all $w$, as we have outlined in Section  \ref{sec:intpro}.
Obviously, we shall obtain very different results than we did there for
flat FLRW data -- depending on the exact character of our more general data.
This completes the framework for solving these exact spherically symmetric
field equations for adequate data when $\Lambda \neq 0$ and the space-time is
not flat. \\ 

\section{\label{sec:concl}Conclusion}

In this paper we have shown in detail how to construct flat dust-filled
$\Lambda \neq 0$ Friedmann-Lema\^{\i}tre-Robertson-Walker cosmological models
from FLRW cosmological data on our past light cone, by integrating
the exact spherically symmetric Einstein field equations in observational
coordinates, extending the approach developed by Ara\'ujo and Stoeger  \cite{AS}
and improved in Ara\'ujo {\it et al}  \cite{ASR}. Besides being of interest in its
own right -- demonstrating how such data fully determines the models --
it also illustrates in a simple case how the more general SS Einstein
equations can be integrated in observational coordinates with data arrayed
on our past light cone, $w=w_0$. Then, we have gone on to show how to
integrate these exact spherically symmetric (LBT) equations, also for
$\Lambda \neq 0$ in cases where the data are not FLRW, and the space-time is
not known to be flat. It is essential for these to have data giving the
maximum of the observer area (angular-diameter) distance, $C_0(w_0, z_{\max})$,
and the redshift $z_{max}$ at which that occurs. This enables the determination
of the vacuum-energy density $\mu_{\Lambda}$, which would otherwise remain
undetermined. \\


\begin{thebibliography}{99}


\bibitem{AS} M.E. Ara\'ujo and W.R. Stoeger,
\textit{Phys. Rev. D {\bf 60}}, 104020 (1999).

\bibitem{ASR} M.E. Ara\'ujo, S.R.M.M. Roveda, and W.R. Stoeger,
\textit{Astrophys. J. {\bf 560}}, 7 (2001).

\bibitem{WMAP} D.N. Spergel et al. 
arxiv: astro-ph/0603449 (2006).

\bibitem{Ellis et al} G.F.R. Ellis, S.D. Nel, R. Maartens, W.R. Stoeger,
and A.P. Whitman,
 \textit{Phys. Rep. {\bf 124}},  315 (1985).

\bibitem{OC III}  W.R. Stoeger, G.F.R. Ellis, and S.D. Nel,
 \textit{Class. Quantum Grav. {\bf 9}}, 509 (1992).
 
\bibitem{SW} R.K. Sachs  and A.M. Wolfe, 
\textit{Astrophys. J.  {\bf 147}}, 73 (1967).

\bibitem{SAG} W.R. Stoeger, M.E. Ara\'ujo,  and T. Gebbie,
\textit{Astrophys. J. {\bf 476}}, 435 (1997).

\bibitem{KS}  J. Kristian and R.K. Sachs, 
\textit{Astrophys. J. {\bf 143}}, 379 (1966).

\bibitem{Temple}  G. Temple,
 \textit{Proc. R. Soc. London A {\bf 168}}, 122 (1938).

\bibitem{Etherington33} I.M.H. Etherington,
 \textit{Philos. Mag. {\bf 15}}, 761 (1933).

\bibitem{Ellis 1971}  G.F.R. Ellis, in 
\textit{General Relativity and Gravitation}, edited by  R.K. Sachs (Academic, New York,1971) p. 104.

\bibitem{Hellaby} C.W. Hellaby,
\textit{Mon. Not. R. Astron. Soc. { \bf 370}}, 239 (2006).

\bibitem{ET}  G.F.R. Ellis and G. Tivon,  
\textit{Observatory  { \bf 105}}, 189 (1985).

\bibitem{ASII} M.E. Ara\'ujo and W.R. Stoeger,
arxiv: astro-ph/0705.1846v2 (2007).

\bibitem{hmm} N. P. Humphreys, D. R. Matravers,  and R. Maartens,
\textit{Class. Quantum Grav. {\bf 15}}, 3041 (1998).

\bibitem{m} R. Maartens, Ph.D. thesis, University of Cape Town, 1980.

\bibitem{fluid ray}  W. R. Stoeger, S.D. Nel,  R. Maartens, and  G.F.R. Ellis,
\textit{Class. Quantum Grav. {\bf 9}}, 493 (1992).

\bibitem{InhomUniv} R. Maartens, N. P. Humphreys, D. R. Matravers,  and
W. R. Stoeger,
\textit{Class. Quantum Grav. {\bf 13}}, 253 (1996); {\bf 13}, 1689 (erratum) (1996).

\bibitem{Lemaitre} G. Lema\^{\i}tre,  
\textit{C. R. Acad. Sci. {\bf 196}}, 903 (1933); {\bf 196} 1085 (1933).

\bibitem{Tolman}  R. C. Tolman,
\textit{Proc. Nat. Acad. Sci. U.S.A.  {\bf 20}}, 169 (1934).

\bibitem{Bondi} H. Bondi,
 \textit{Mon. Not. R. Astron. Soc. { \bf 107}}, 410 (1947).

\bibitem{HT} H. P. Humphreys, Ph.D. thesis, University of Portsmouth 1998.

\bibitem{Bonnor}  W. B. Bonnor, 
\textit{Mon. Not. R. Astron. Soc. { \bf 167}}, 55 (1974).

\bibitem {ES1987} G.F.R. Ellis and W. R. Stoeger,
\textit{Class. Quantum Grav. {\bf 4}}, 1697 (1987).

\end{thebibliography}
\end{document}